\documentclass{emulateapj}

\def\etal{{et~al.\,}}

\def\sles{\lower2pt\hbox{$\buildrel {\scriptstyle <}
   \over {\scriptstyle\sim}$}}
\def\sgreat{\lower2pt\hbox{$\buildrel {\scriptstyle >}
   \over {\scriptstyle\sim}$}}

\slugcomment{Accepted Aug. 28, 2007 to Ap.J. Letters}

\begin{document}

\title{Theoretical Spectral Models of the Planet HD 209458b with a 
Thermal Inversion and Water Emission Bands}

\author{A. Burrows\altaffilmark{1}, I. Hubeny\altaffilmark{1}, J. Budaj\altaffilmark{1,2}, 
H.A. Knutson\altaffilmark{3}, \& D. Charbonneau\altaffilmark{3}} 

\altaffiltext{1}{Department of Astronomy and Steward Observatory, 
                 The University of Arizona, Tucson, AZ \ 85721;
                 burrows@zenith.as.arizona.edu, hubeny@aegis.as.arizona.edu, budaj@as.arizona.edu}

\altaffiltext{2}{Astronomical Institute, Tatranska Lomnica, 05960 Slovak Republic} 

\altaffiltext{3}{Harvard-Smithsonian Center for Astrophysics, 60 Garden Street, 02138 USA; 
                 hknutson@cfa.harvard.edu, dcharbonneau@cfa.harvard.edu} 

\begin{abstract}

We find that a theoretical fit to all the HD 209458b data at secondary eclipse
requires that the dayside atmosphere of HD 209458b have a thermal inversion 
and a stratosphere.  This inversion is caused by the capture of optical
stellar flux by an absorber of uncertain origin that resides at altitude.
One consequence of stratospheric heating and temperature inversion is the flipping 
of water absorption features into {\it emission} features from the near- to the mid-infrared
and we see evidence of such a water emission feature in the recent HD 209458b IRAC data 
of Knutson et al. In addition, an upper-atmosphere optical absorber may help explain both
the weaker-than-expected Na D feature seen in transit and the fact that the 
transit radius at 24 $\mu$m is smaller than the corresponding radius 
in the optical. Moreover, it may be a factor in why HD 209458b's optical transit radius 
is as large as it is. We speculate on the nature of this absorber and the planets
whose atmospheres may, or may not, be affected by its presence. 

\end{abstract}


\section{Introduction}
\label{intro}

Of the more than 235 extrasolar giant planets (``EGPs") discovered in 
the last twelve years \footnote{See J. Schneider's Extrasolar Planet 
Encyclopaedia at http://exoplanet.eu, the Geneva Search Programme at
http://exoplanets.eu, and the Carnegie/California compilation at http://exoplanets.org}, 22 are
transiting their primaries.  The most thoroughly studied transiting EGP 
is HD 209458b (Henry et al. 2000; Charbonneau et al. 2000; Brown et al. 2001;
Melo et al. 2006; Santos, Israelian, \& Mayor 2004; Knutson et al. 2007a). 
For transiting EGPs, not only do we measure the planet's radius, 
but the $\sin{i}$ ambiguity of radial-velocity studies is resolved to yield the 
planet's mass. Moreover, with precision photometry, the wavelength-dependence 
of the transit radii can provide a measure of a planet's atmospheric 
composition (Burrows et al. 2000; Hubbard et al. 2001; 
Fortney et al. 2003; Barman 2007).  In this way, sodium (Charbonneau et al. 2002)
and water (Barman 2007) have been identified in HD 209458b and 
water has been identified in HD 189733b (Tinetti et al. 2007)
and in TrES-1 (Burrows, Hubeny, \& Sudarsky 2005). 
Moreover, using the micro-satellite MOST, 
high-precision optical photometry has constrained (perhaps, measured) HD 209458b's geometric albedo
(Rowe et al. 2006, 2007).  It is very low ($\sim$4.0$\pm4.0$\%), 
in keeping with the predictions of Sudarsky, Burrows, 
\& Pinto (2000) when the alkali metals, and not clouds, dominate absorption in 
the atmosphere and Rayleigh scattering dominates scattering.  
This makes HD 209458b darker in the optical than most
of the objects of our solar system and also suggests that 
the associated contrast ratios are not optimal for 
characterizing EGPs. 

However, for hot, close-in EGPs such as HD 209458b, the 
planet-star contrast ratios in the mid-infrared are much more favorable 
than in the optical (Burrows, Sudarsky, \& Hubeny 2003; Burrows, Sudarsky, \& Hubeny 2004; 
Burrows 2005), oftimes exceeding 10$^{-3}$, and such contrasts are within reach of 
the infrared space telescope {\it Spitzer} (Werner \& Fanson 1995).  
Using its IRAC and MIPS cameras and the IRS spectrometer, one can now measure the 
summed light of the planet and the star in and out of secondary eclipse 
(when the star occults the planet) and, from the difference, determine 
the planet's spectrum at superior conjunction.  
This has led to a breakthrough in the study of exoplanets
and a means to probe their chemistry and atmospheres.
Hence, for the close-in EGPs in the near- to mid-infrared, and without
the need to separately image planet and star, the direct detection of
planetary atmospheres via low-resolution spectroscopy and precision IR photometry
is now a reality (Charbonneau et al. 2005; Deming et al. 2005,2006,2007; Grillmair et al. 2007; 
Richardson et al. 2007; Harrington et al. 2006,2007; Knutson et al. 2007b,c; 
Cowan et al. 2007).  

To date, secondary eclipse fluxes in the IRAC and MIPS channels have been measured 
for five transiting EGPs (HD 189733b, TrES-1, HD 209458b, HD 149026b, and GJ 436b), but in 
this paper we focus on the interpretation of the HD 209458b data. A more 
comprehensive paper on our numerical technique and theoretical fits 
to other secondary eclipse data, as well as to phase light curve measurements, 
is in preparation (Burrows, Budaj, \& Hubeny 2007b).  For HD 209458b, not only is there 
a complete set of measurements at secondary eclipse using IRAC and 
MIPS, but using the IRS spectrometer, a low-resolution spectrum between 
$\sim$7.5 $\mu$m and $\sim$15 $\mu$m has now been obtained (Richardson et al. 2007; 
Swain et al. 2007). 

Excitingly, the new IRAC data of Knutson et al. (2007c) trace out a positive 
bump in the $\sim$3.6$\mu$m to 8 $\mu$m spectral region, which we interpret via detailed modeling as a  
water {\it emission} feature.  A water absorption feature was anticipated (Burrows, 
Hubeny, \& Sudarsky 2005; Burrows, Sudarsky, \& Hubeny 2006; Fortney et al. 2005;
Barman, Hauschildt, \& Allard 2005).  Therefore, this is strong evidence for a thermal inversion in the atmosphere of
HD 209458b which has flipped water absorption features into emission features.  We now provide
the comparison between theory and the observations and the resulting preliminary analysis.

\section{Fit to the HD 209458b Planet/Star Flux Ratios at Secondary Eclipse}

Three models with an upper-atmosphere absorber in the optical (and with 
the resulting stratospheres) for the planet/star flux ratios of HD 209458b 
at secondary eclipse are portrayed in Fig. \ref{fig1}.  The purple, green,
and red models are for redistribution parameters (P$_n$; Burrows, Sudarsky, 
\& Hubeny 2006) of 0.3, 0.4, and 0.5, respectively, but are otherwise
the same.  For each model, an extra absorber of uncertain provenance
is placed at altitude below pressures of 25 mbars.  The total monochromatic 
optical depth of this layer is $\sim$3. A new redistribution algorithm that introduces a heat sink 
on the dayside between 0.01 and 0.1 bars, and a corresponding source on the nightside, is 
employed. This algorithm is explained in Burrows, Budaj, \& Hubeny (2007b). 
Following the suggestion of Hubeny, Burrows, \& Sudarsky (2003) in 
their atmosphere bifurcation study, we have also calculated a 
model suite (not shown) with equilibrium TiO and VO in the stratosphere.
With a total Planck-mean optical depth of TiO/VO below $\sim$ 30 mbars 
of $\sim$3.1, these models are quite similar to those with inversions 
depicted in Fig. \ref{fig1} with the same P$_n$s. Figure \ref{fig1} also 
provides a representative default model (black) without a stratospheric 
%
%
absorber, but with P$_n$ = 0.3, 
along with all the relevant HD 209458b measurements to date.  The 
default model represents the previous predictions for atmospheres with 
monotonically decreasing temperatures and no inversions, though the new 
redistribution algorithm alluded to above was employed.  

The most salient observational constraints for our current purposes are the geometric 
albedo in the optical from MOST (Rowe et al. 2006,2007), a K-band upper limit
using IRTF/SpeX from Richardson, Deming, \& Seager (2003), a MIPS/24-$\mu$m photometric point 
from Deming et al. (2005), a low-resolution {\it Spitzer}/IRS spectrum from 
Richardson et al. (2007), and, most importantly, photometric points 
in IRAC channels 1 through 4 from Knutson et al. (2007c). Richardson et al. (2007) 
suggest that there is evidence in the IRS data for two spectral features: one near
7.78 $\mu$m and one near 9.67 $\mu$m.  However, we think the data are too noisy to
draw this conclusion and await the next generation of observations to test it.
Moreover, we note that Richardson et al. (2007) normalize their data to the Knutson et 
al. (2007c) IRAC 4 point, but that due to the noisiness of the IRS data near 
8 $\mu$m this normalization deserves a second look.

The 1-$\sigma$ optical albedo limit from 
Rowe et al. (2007) is $\sim$8.0\%, a very low number.  For comparison, the 
geometric albedo for Jupiter is $\sim$40\%.  However, such a low number was predicted
due to the prominence in the optical of broadband absorptions by the alkali metals sodium 
and potassium in the hot atmospheres of irradiated EGPs (Sudarsky et al. 2000; 
their ``Class IV").  Alkali metals are not found in Jupiter's atmosphere.  
The associated planet-star flux ratios are $\sim$10$^{-5}$$-$10$^{-6}$.  
A low albedo due to absorption by the Na D and K I resonance lines is consistent with the identification of sodium 
in the atmosphere of HD 209458b using HST/STIS transit spectroscopy (Charbonneau 
et al. 2002), though the magnitude of the sodium feature has yet to be fully explained.
Both these datasets suggest that any clouds that might reside in the 
atmosphere of HD 209458b are thin or mostly absorbing.  A thick scattering cloud layer 
would reflect light rather efficiently and lead to a high albedo, which is not seen.
The extra absorber we introduce could in fact be a cloud, but in that case it must be comprised of particles
with a low scattering albedo.  This would seem to eliminate forsterite and enstatite.

As Fig. \ref{fig1} demonstrates, the low upper limit of Richardson, Deming, \& Seager (2003) in the K band
that was problematic in the previous theory (Burrows, Hubeny, \& Sudarsky 2005; Fortney et al. 2005;
Barman, Hauschildt, \& Allard 2005) is comfortably consistent with the models with an extra
upper-atmosphere absorber in the optical, particularly for higher values of P$_n$.
Moreover, the peak near the IRAC 1 channel ($\sim$3.6 $\mu$m) in the previous 
model without an inversion (Burrows et al. 2006) is reversed with the extra absorber into a deficit
that fits the Knutson et al. (2007c) point. Importantly, the theory without an extra absorber
at altitude predicts that the planet-star flux ratio in the IRAC 2 channel should be lower than
the corresponding ratio at IRAC 1.  However, with the extra absorber the relative strengths in
these bands are reversed, just as are the Knutson et al. (2007c) points for HD 209458b.  This reversal
is a clear signature of a thermal inversion in the low-pressure regions of the atmosphere. 
Figure \ref{fig2} depicts the corresponding temperature-pressure profiles and the thermal inversion
at low pressures introduced by the presence of an extra absorber in the optical. 
It also indicates the approximate positions of the effective photospheres for photons 
in the IRAC and MIPS channels.  The fact that in the new models the IRAC 1 photosphere 
is cooler than the IRAC 2 photosphere is a key to the observed behavior of the Knutson et al. (2007c) data.

Fig. \ref{fig1} indicates that the models with a stratosphere fit the IRAC 
1, 2 and 4 flux points quite well.  However, the relative height of the 
IRAC 3 point near 5.8 $\mu$m is difficult to fit, while maintaining the good
fits at the other IRAC points and consistency with the Richardson, Deming, \& Seager 
(2003) limit. As Fig. \ref{fig2} shows, the 
positions of the IRAC 3 and IRAC 4 photospheres are generally close to one 
another. This makes it difficult to have effective ``brightness" 
temperatures that are as different as are implied by the IRAC data. 
Nevertheless, IRAC 2, 3, and 4 together trace out a peak, whereas in 
the default theory an absorption trough was expected.  Note in 
Fig. \ref{fig1} the significant difference in the planet-star flux ratio 
at these wavelengths between the old default model and the new models with 
a stratospheric absorber. This is the spectral region of a strong 
ro-vibrational band of water and a comparison between our new HD 209458b 
models and the data indicates that this feature is in emission.  

The IRS data of Richardson et al. (2007) are normalized to the Knutson et al. (2007c) IRAC 4 point, 
so the good fit there is not independent.  Nevertheless, the IRS data have a flattish slope 
between $\sim$7.5 $\mu$m and $\sim$14 $\mu$m that is mildly inconsistent with the slight rise 
of our new models.  In addition, the 24-$\mu$m MIPS point obtained by 
Deming et al. (2005) is lower than these models.  However, the flux at 
this point is being reevaluated (D. Deming, private communication). If
the new value at 24 $\mu$m is, as suggested, $\sim$0.0033$\pm{0.0003}$, then our new model(s) with 
inversions fit here as well (see Fig. \ref{fig1})\footnote{Note, though, that the actual updated number 
has yet to be determined and published.}. Be that as it may, the good qualitative and quantitative fits
for the K band limit, three of the four IRAC channels, and the MOST albedo limit (not shown) lend credence to
our overall model and conclusions.  Concerning the dayside atmosphere of HD 209458b, there is an extra absorber 
in the optical at altitude, there is a pronounced thermal inversion (see Fig. \ref{fig2}), 
and water is seen in emission.

\section{Discussion}
\label{conclusions}

\begin{figure}
\centerline{
\includegraphics[width=14.cm,angle=-90,clip=]{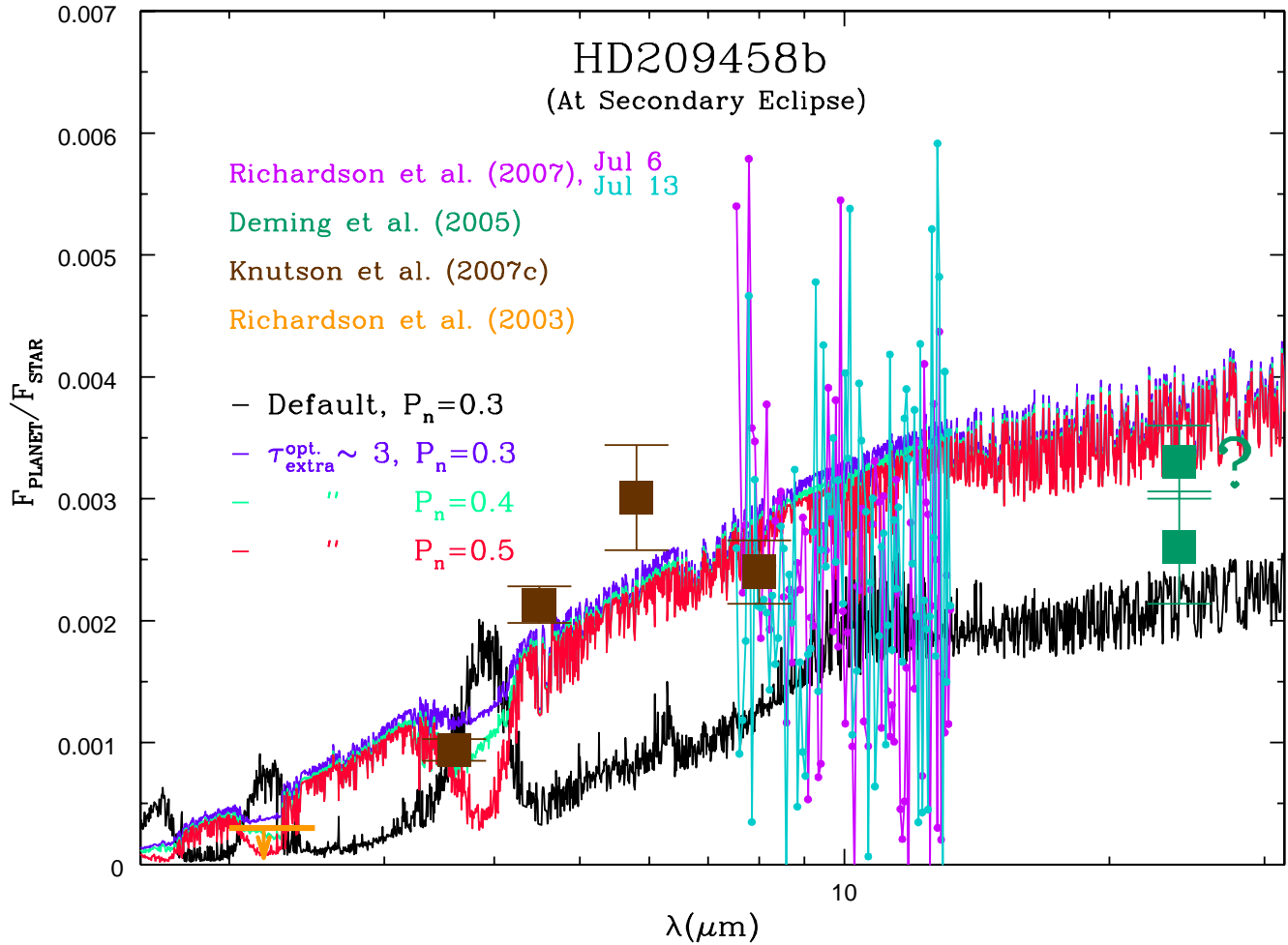}}
\caption{
The planet-star flux ratios at secondary eclipse versus wavelength for 
three models of the atmosphere of HD 209458b with inversions and for one model without an 
extra upper-atmosphere absorber of any kind (black, old default).
The three models with stratospheres have different values of P$_n$ (= 0.3 [purple],
0.4 [green], and 0.5 [red]), but are otherwise the same.  The old, default model
has a P$_n$ of 0.3.  This figure demonstrates that models 
with an extra upper-atmosphere absorber in the optical and with 
P$_n \sgreat 0.35$ fit the data; the old, default model fits not at all.  
We have also calculated models (not shown) with equilibrium TiO/VO in the 
upper atmosphere ($\le$ 30 mbars) (Burrows, Sudarsky, \& Hubeny 
2006; Burrows, Budaj, \& Hubeny 2007b) which are quite 
close to the corresponding models with an extra absorber. 
Superposed are the data in the K band ($\sim$2.2 $\mu$m) from Richardson, 
Deming, \& Seager (2003) (orange line and arrow), the four IRAC points 
from Knutson et al. (2007c) (brown square blocks), the IRS spectra from Richardson et 
al. (2007) (purple and aqua), and the MIPS data at 24 $\mu$m from Deming et al. (2005) (green square block).  
Also included, with a question mark beside it, is a tentative update to this 24 $\mu$m flux point,
kindly provided by Drake Deming (private communication). If the flux at 
24 $\mu$m is indeed $\sim$0.0033$\pm{0.0003}$, then our model(s) 
with inversions fit there as well.  See the text for discussions.
}
\label{fig1}
\end{figure}

\begin{figure}
\centerline{
\includegraphics[width=14.cm,angle=-90,clip=]{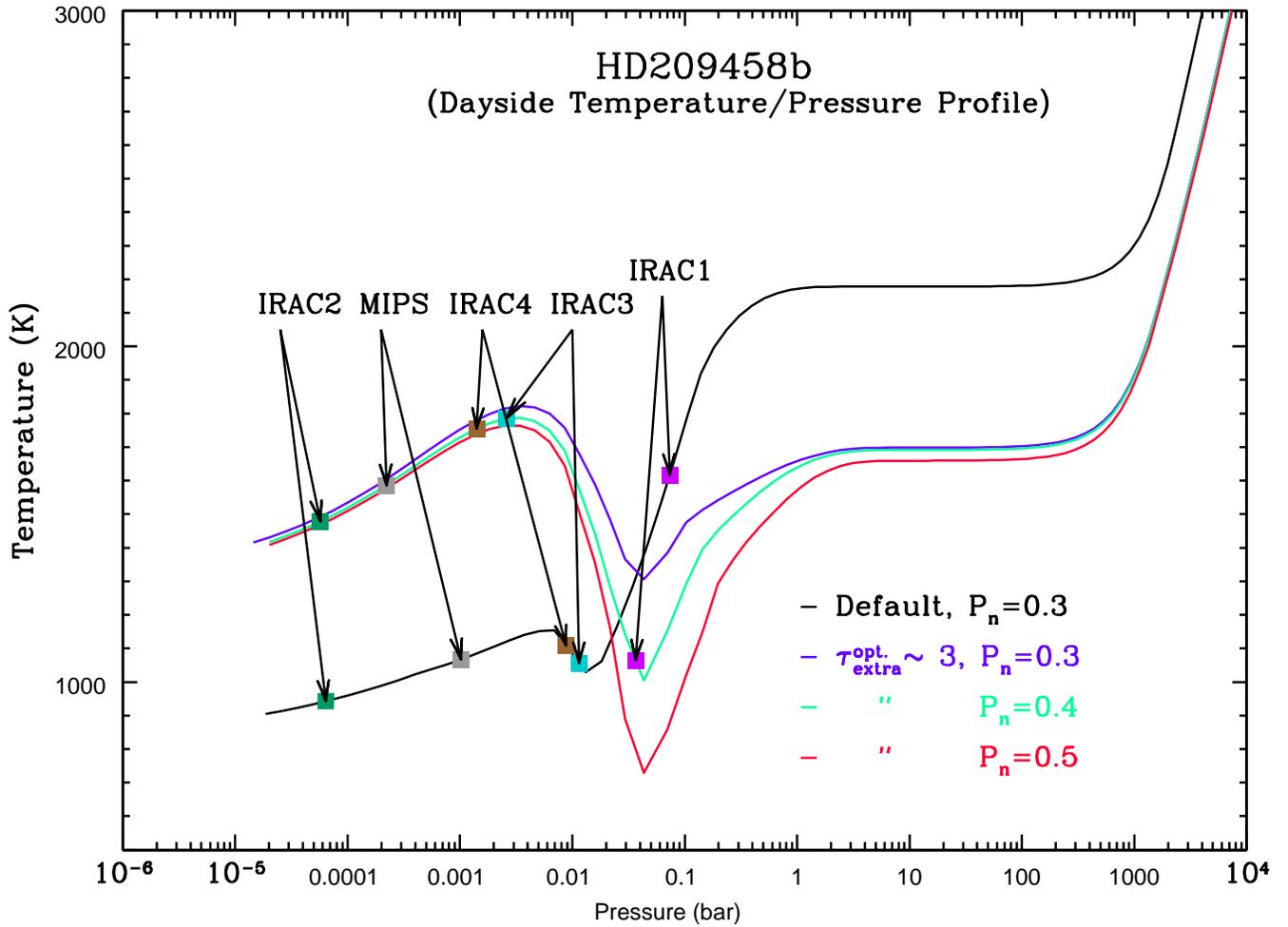}}
\caption{
Dayside temperature (K)-pressure (bars) profiles for the four models for HD 209458b depicted in Fig. \ref{fig1}.
For the old, default model and the new P$_n$ = 0.4 model, we indicate the positions of 
the ``photospheres" for the IRAC and MIPS/24-$\mu$m bands, defined as the corresponding 
$\tau =2/3$ surfaces. Most of these mid-infrared photospheres are at altitude, where 
an extra absorber can have a significant effect.  The photospheres for the near-IR J, H, 
and K bands (as well as the IRAC 1 band) are deeper in.  A comparison of the green and the black curves
demonstrates that both cooling by heat redistribution (P$_n \sgreat 0.3$) and heating by absorption
in the upper atmosphere together are necessary to invert the IRAC1/IRAC2 flux ratio.
All three inversion models experience some cooling at moderate pressures 
due to a sink imposed to mimic heat redistribution, but this process alone is 
insufficient to elevate the temperatures of the upper atmosphere to values necessary 
to fit the Knutson et al. (2007c) IRAC data for HD 209458b. An extra absorber in the 
optical and at low pressures is called for.  See text for a discussion. 
}
\label{fig2}
\end{figure}

We find that a consistent fit to all the HD 209458b data at secondary eclipse
requires that the atmosphere of HD 209458b have a thermal inversion at altitude
and a stratosphere.  This inversion is caused by the capture of incident optical 
stellar flux by an extra absorber of currently uncertain origin that resides at 
low pressures.  The IRAC data of Knutson et al. (2007c) can not be
fit by the effects of dayside heat redistribution cooling alone.  A consequence
of stratospheric heating and temperature inversion is the flipping of water absorption features into emission
features from the near- to the mid-infrared (Hubeny, Burrows, \& Sudarsky 2003; Burrows, 
Sudarsky, \& Hubeny 2006), and this seems to explain all the current HD 209458b data.  Hence, 
contrary to the interpretation of Richardson et al. (2007), the flatness or slight rise
of their IRS spectrum near $\sim$7.8 $\mu$m (Fig. \ref{fig1}) in fact supports the presence of 
abundant atmospheric water, because this region is at the edge of a strong water band in emission.
Our inference of the presence of water in abundance in the atmosphere of HD 209458b is also consistent with the
conclusion of Barman (2007) using transmission spectroscopy.

If the extra absorber is in the gas phase, and there is no cloud, then 
our new models are easily consistent with the low albedo derived by Rowe et al. (2006,2007).
If the extra absorber is a cloud, the cloud particles must have a low scattering albedo
and can not be very reflecting.  This would seem to rule out forsterite or enstatite
clouds, but does not necessarily rule out iron clouds.  As shown by Hubeny, Burrows, \&
Sudarsky (2003) and Burrows, Sudarsky, \& Hubeny (2006), strongly irradiated atmospheres
can experience a solution bifurcation to an atmosphere with an inversion for which 
the water features are flipped from absorption to emission.  In
those papers, the absorber was gas-phase TiO/VO, which in equilibrium can exist at low pressures
at altitude and not just at high temperatures at depth.  If the extra absorber were
TiO/VO, there would be fewer free parameters, but it was thought that the ``cold-trap"
effect would quickly deplete the upper atmosphere of TiO/VO and ensure the default
atmospheric solution without an inversion.  However, this has not been proven, particularly
when atmospheric mass loss is ongoing, as we know to be the case for HD 209458b 
(Vidal-Madjar et al. 2003,2004).  The same arguments hold for iron clouds, although
since absorption by iron particles is not restricted to the optical,
this solution is sub-optimal.   

One can also speculate that the severe irradiation regime of some close-in EGPs
might create non-equilibrium compounds through photolysis, such as the tholins or
polyacetylenes discussed in the more benign contexts of solar-system bodies,
that could serve as the extra absorber we deduce exists in the atmosphere 
of HD 209458b.  Clearly, what the high-altitude absorber actually is remains to be seen,
but one can speculate that its presence is more likely in the atmospheres
of the most strongly irradiated EGPs.  What ``most strongly irradiated" actually
means is not yet clear, but the flux at the substellar point on HD 209458b is $\sim$$10^{9}$ erg cm$^{-2}$ s$^{-1}$.
The corresponding numbers for OGLE-TR-10b, OGLE-TR-56b, OGLE-TR-132b, TrES-2, TrES-3,
WASP-1b, XO-3b, HAT-P-1b, and HD 149026b are higher (Burrows et al. 2007a).  The corresponding number for 
TrES-1 is lower ($\sim$0.43$\times10^{9}$ erg cm$^{-2}$ s$^{-1}$) and this planet shows good evidence 
for water in {\it absorption} (Burrows, Hubeny, \& Sudarsky 2005).  In addition, the
IRS spectrum of HD 189733b of Grillmair et al. (2007) seems consistent with a more canonical
water absorption feature shortward of $\sim$8 $\mu$m.  Its substellar flux is 
$\sim$0.47$\times10^{9}$ erg cm$^{-2}$ s$^{-1}$.  So, if stellar flux at
the planet is an indicator, we may have a handle on which planets reside in the transition 
region between manifesting water absorption or emission features (and inversions), and where the 
relative IRAC 1/IRAC 2 strengths start to flip.  Given their substellar fluxes, HD 189733b, 
XO-1b, XO-2b, and/or WASP-2b may be links.  Though we find a weak dependence 
on metallicity, non-equilibrium chemistry and cloud formation may have stronger dependences.  
Therefore, there are still numerous parameters to address.

As Fig. \ref{fig1} suggests, the planet/star ratios shortward of $\sim$4 $\mu$m are
more sensitive to the heat redistribution parameter, P$_n$, than the corresponding
ratios longward of $\sim$8 $\mu$m.  This suggests a shorter-wavelength strategy 
for designing close-in EGP climate diagnostics. In addition, models with inversions 
significantly boost the mid-infrared fluxes longward of $\sim$15 $\mu$m. This boost 
is not the only signature of models with dayside stratospheres.  Since the 
nightside flux at the same P$_n$ is unlikely to be much altered by the fact 
of dayside inversion, if the dayside has such an inversion due to enhanced 
optical absorption at altitude, the day-night contrasts measured during an orbital
traverse by a close-in EGP will be larger than expected for a given P$_n$. This 
possibility is relevant when interpreting the 24-$\mu$m light curve of $\upsilon$ And b
(Harrington et al. 2006).  Perhaps, the large day-night contrast seen in this 
case does not require a small P$_n$.

An upper-atmosphere absorber in the optical may simultaneously explain
the weaker-than-expected Na D feature seen in transit by Charbonneau et al. (2002)
(Fortney et al. 2003) and the fact that the transit radius measured by Richardson et al. (2006) at 24 $\mu$m
(1.26$\pm$0.08 R$_{\rm J}$) is smaller than the corresponding radius in the optical
(1.32$\pm$0.025 R$_{\rm J}$; Knutson et al. 2007a).  In addition, it may help explain 
why HD 209458b's optical transit radius is as large as it is.  Finally, the extra 
absorber may be related to the enhanced opacities for close-in EGPs discussed 
in the context of the EGP radius models of Burrows et al. (2007a).  Whatever their 
actual roles, anomalous optical absorption at altitude and thermal inversions 
are now insinuating themselves as exciting new components of EGP theory.

\acknowledgments

We thank Drake Deming, Bill Hubbard, Maki 
Hattori, Mike Cushing, and Drew Milsom for helpful discussions.
This study was supported in part by NASA grants NNG04GL22G and NNX07AG80G and 
through the NASA Astrobiology Institute under Cooperative 
Agreement No. CAN-02-OSS-02 issued through the Office of Space
Science.


{}

\clearpage

\clearpage

\end{document}